\documentclass[12pt,preprint]{aastex}

\begin{document}

\title{Gas phase water in the surface layer of protoplanetary disks}

\author{
C. Dominik\footnote{Astronomical Inst. ``Anton Pannekoek'', Univ. of
 Amsterdam, Kruislaan 403, NL-1098SJ Amsterdam : dominik@science.uva.nl}, 
C. Ceccarelli\footnote{Laboratoire d'Astrophysique, Observatoire de Grenoble -
 BP 53, F-38041 Grenoble cedex 09, France : 
 Cecilia.Ceccarelli@obs.ujf-grenoble.fr},
D.Hollenbach\footnote{NASA Ames Research Center, Moffett Field, CA
  94035 : hollenbach@ism.arc.nasa.gov},
M.Kaufman$^{3,}$\footnote{Dept. of Physics, San Jose State University,
  One Washington Square, San Jose, CA 95192-0106 : kaufman@ism.arc.nasa.gov}
}

\begin{abstract}
  Recent observations of the ground state transition of HDO at 464 GHz
  towards the protoplanetary disk of DM Tau have detected the presence
  of water vapor in the regions just above the outer disk midplane
  \citep{HDO-DMTAU}.  In the absence of non-thermal desorption
  processes, water should be almost entirely frozen onto the grain
  mantles and HDO undetectable. In this Letter we present a chemical
  model that explores the possibility that the icy mantles are
  photo-desorbed by FUV (6\,eV $\le h\nu \le$ 13.6\,eV) photons.  We
  show that the average Interstellar FUV field is enough to create a
  layer of water vapor above the disk midplane over the entire disk.
  Assuming a photo-desorption yield of $10^{-3}$, the water
  abundance in this layer is predicted to be $\sim 3\times 10^{-7}$
  and the average H$_2$O column density is $\sim 1.6\times 10^{15}$
  cm$^{-2}$.  The predictions are very weakly dependent on
    the details of the model, like the incident FUV radiation field,
    and the gas density in the disk.  Based on this model, we predict
  a gaseous HDO/H$_2$O ratio in DM Tau of $\sim1$\%.  In addition, we
  predict the ground state transition of water at 557 GHz to be
  undetectable with ODIN and/or HSO-HIFI.
 \end{abstract}

\def\fdep{\ensuremath{f_\mathrm{dep}}}
\def\kf{\ensuremath{k_\mathrm{f}}}

\keywords{Stars: formation -- Stars: protoplanetary disks -- Stars: Pre-main-sequence
-- ISM: molecules -- }

\setcounter{footnote}{0}

\section{Introduction}

Water is a key ingredient in many astrophysical environments,
including circumstellar disks. It is the main constituent of the icy
mantles coating interstellar grains, and can be one of the most
abundant molecules in the gas phase. For these reasons, water may
dominate both the chemistry (being a major oxygen reservoir) and the
thermal balance (i.e. cooling) of the gas component of protostellar
disks.  In addition, water vapor and water ice in protoplanetary
disks are the major reservoirs of water later to be found in planets,
asteroids and comets and, therefore, are of great relevance for
understanding the origin of the Solar System and the distribution of
volatiles within it.  Unfortunately, the observation of water
molecules in interstellar space is nearly impossible from the ground
because of the absorption by the terrestrial atmosphere.  So far, just
a handful of space-based instruments (ISO, SWAS and ODIN) have carried
out observations of water vapor lines in interstellar molecular
clouds, but these instruments did not have the sensitivity to observe
water protoplanetary disks. However, the up-coming ESA/NASA mission
Herschel will provide a sensitive instrument, HIFI, for these studies
\citep{2005dmu..conf..171D}.  Also, SOFIA has potential for water
observations.

Meanwhile, HDO, as the most important isotope of water, is currently
the best ground-based probe of the presence of water in astrophysical
environments.  Recently, \citet{HDO-DMTAU} reported the first
discovery of HDO (and therefore also the first discovery of water) in
a protoplanetary disk, namely DM Tau.  The observations show an
absorption line of gas phase HDO with a line-to-continuum ratio of 0.9
which translates into an average column density through the disk of
$\sim 1.6\times 10^{13}$ cm$^{-2}$.  In the region where HDO is
  located, its fractional abundance is estimated to be about $3\times
  10^{-9}$. If the abundance of gaseous H$_2$O is similar to the value
  observed in molecular clouds and protostars, $\sim 10^{-7}$,
  HDO/H$_2$O ratio is 0.01.
The fact that the line is a deep absorption line immediately indicates
that HDO must be present above the midplane (from where the
sub-millimeter continuum is emitted).  It must also be present
especially in the outer radial regions ($\geq 600$ AU) of the disk
since the continuum originates from an extended region.  The presence
of a large column of HDO in this location was a surprise,
since the grain temperatures in this region are below 25 K, and H$_2$O
and HDO should be almost entirely frozen out onto grains.
Chemical models including \emph{thermal} desorption of water
  ice and gas-phase formation of H$_2$O
  \citep{2002A&A...386..622A,2003A&A...397..789V} predict a H$_2$O
  column of 10$^{14}$ cm$^{-2}$ at about 400\,AU, only a factor of 5
  larger than the observed HDO column.  \citet{2000ApJ...544..903W}
  included photo-desorption by a strong stellar UV field and found
  H$_2$O column densities of about $5\times10^{15}$cm$^{-2}$ at
  700\,AU.
An extended component of HDO (and consequently
water) therefore indicates a desorption agent from grain surfaces.
Important possibilities are X-rays originating from the star, and
penetrating throughout the entire outer disk, and UV photons, either
from the star or simply due to the interstellar radiation field.

High energy photons can in principle act to remove H$_2$O molecules
from the ice layers on dust grains.  X-rays are energetic enough to
remove a number of molecules, if sufficient energy can be concentrated
close to the surface of the ice mantle on a grain.  This
idea has been explored by \citet{2001ApJ...561..880N} who considered X-ray
heating of small grains, or spot heating of larger grains.  Small
grains, or small thermally insulated spots on large grain can be
heated strongly enough by a single X-ray absorbed photon to lead to thermal
evaporation of part of the ice mantle.

An alternative mechanism is desorption by FUV photons
\citep{2000ApJ...544..903W}.  An FUV photon absorbed in the surface
layer of an ice mantle can lead to the release of a molecule into the
gas.  In this Letter we explore the effect of just the interstellar
FUV radiation field irradiating the disk surface of DM Tau.
We show that such a scenario leads to a layer of water vapor in
the disk surface to an FUV optical depth around unity.  The column
densities reached in steady state appear to be a viable explanation
for the HDO absorption line seen in DM Tau.  To show this we developed a
simple model of the chemistry leading to the presence of water vapor
in irradiated gas containing grains covered by layers of water ice.
A much more detailed discussion in the context of a full
Photo-Dissociation Region (PDR) model, as well as analytical formulas
to estimate H$_2$O column densities under such conditions will be
given by Hollenbach et al (2005, in preparation).  Here we report the
application of this model to the case of DM Tau. A forthcoming paper
will present the results of a larger parameter space study, which will
explore the dependence of the model predictions especially on the
characteristics of the grains (Ceccarelli et al. in preparation).

\section{Model description}

\begin{figure}[htbp]
\centerline{\includegraphics[width=9.0cm, angle=90]{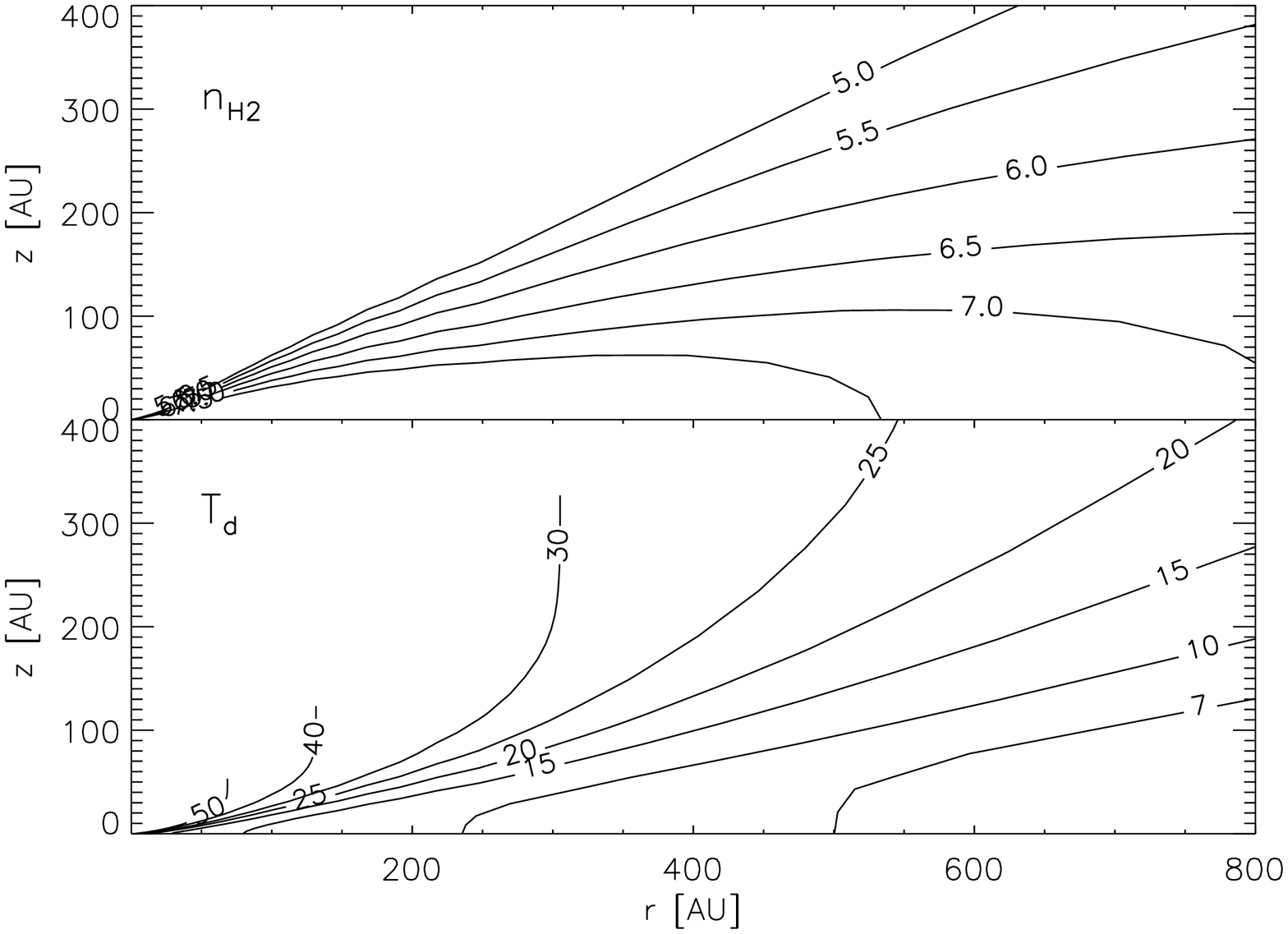}}
\caption{The logarithm of the gas density (upper panel) and the
  temperature (lower panel) of the disk surrounding DM Tau, as derived
  by modeling of the SED \citep{HDO-DMTAU}, and assuming that dust and
  gas are coupled with $\rho_\mathrm{dust}/\rho_\mathrm{gas}=0.01$.}
\label{fig:structure}
\end{figure}

The model computes the H$_2$O abundance across a protoplanetary disk,
as a function of the radius and height.  For the physical structure we
used the dust density and temperature profile which fits the Spectral
Energy Distribution (SED) of DM Tau \citep{HDO-DMTAU}, and
shown in Fig. \ref{fig:structure}.  The gas is assumed to be fully
coupled with the dust in terms of density distribution and
temperature.

Due to the low temperature, the dust grains are covered with an ice
layer.  In our model, UV photons entering the disk surface are
absorbed by dust grains, and cause photo-desorption of water molecules
from the ice.  Only UV photons absorbed directly at the surface of the
ice mantle can desorb a molecule, so that only a small fraction of
absorbed photons will cause the ejection of H$_2$O.
Laboratory experiments at Lyman $\alpha$ wavelength have shown that
the involved yields are typically between 10$^{-3}$ and 10$^{-2}$
molecules per photon \citep{1995P&SS...43.1311W}.  The yield increases
to some extent depending on the UV dose in the experiment, apparently
because radical formation on the ice surface aids the desorption
process \citep{1995Natur.373..405W}.  It is unclear how important
this effect is in astrophysical environments, and we adopt a
conservative value for the yield of 10$^{-3}$ molecules per incident
FUV (6$\leq h\nu \leq$13.6 eV) photon.  The yield probably introduce a
factor of a few uncertainty into the computations (see below).

The desorption process at a given location will also depend on the
attenuated FUV flux, and the rate per unit volume can be written as:
\begin{equation}\label{eq:1}
k_\mathrm{des}=G_0 f_0 Y {\rm exp}(-N(H_2)/N_\mathrm{uv}) \sigma_\mathrm{gr} 
   \mathrm{n_{gr}}
\end{equation}
where $G_0$ is the FUV field in \citet{1968BAN....19..421H} units.
$f_0=10^8$ photons/cm$^2$/s is the FUV flux for the standard Habing
interstellar field ($G_0$=1).  $Y$ is the photo-desorption yield.
$N_\mathrm{uv}$ is H$_2$ column density that gives
$\tau_\mathrm{uv}=1$ between 6 and 13.6 eV; we adopted a value equal
to $1.8\times10^{21}$ cm$^{-2}$ \citep{1985ApJ...291..722T}.
$\sigma_\mathrm{gr}$ is the grain geometrical cross section, equal to
$\pi~ a_\mathrm{gr}^2$, where we assumed an average grain size of 0.1
$\mu$m. $n_\mathrm{gr}$ is the grain number density.  The product
$\sigma_\mathrm{gr} n_\mathrm{gr}$ is approximatively $10^{-21}$
cm$^{-2}$ times the gas density $\mathrm{n_{H_2}}$, obtained assuming
a gas to dust ratio in mass equal to 1:100.
After being released into the gas phase, water either freezes out back
onto a grain, or it is photo-dissociated by the FUV photons at the
rates:
\begin{equation}\label{eq:2}
k_\mathrm{freeze}=S_\mathrm{gr} \sigma_\mathrm{gr}^2 \mathrm{n_{gr}}  
 \mathrm{n_{H_2O}} <v_\mathrm{th}>
\end{equation}
and
\begin{equation}\label{eq:3}
k_\mathrm{phd} = G_0 I_0 \exp(-N(H_2)/N_\mathrm{uv})\mathrm{n_{H_2O}} 
\end{equation}
where $S_\mathrm{gr}$ is the sticking coefficient, for which we use 1
\citep{1983ApJ...265..223B}.  $<v_\mathrm{th}>$ is H$_2$O molecule
thermal velocity.  $I_0=5.1\times 10^{-10}$s$^{-1}$
\citep{2000A&AS..146..157L} is the rate of FUV dissociation in
unshielded $G_0=1$ field.

Gaseous atomic oxygen can freeze out onto grains, where it is assumed
to form water ice by reactions with hydrogen atoms on the grain surfaces.
O atoms may also form water in the gas phase via the standard sequence
of reactions started by the reaction between O and H$_3^+$. The gas
phase formation of water thus proceeds at a rate:
\begin{equation}\label{eq:4}
k_\mathrm{gas}=k_\mathrm{form} n_O n_{H_3^+}
\end{equation}
where $k_\mathrm{form}$ is equal to $8\times 10^{-10}$ s$^{-1}$
cm$^{-3}$ (the rate coefficient for the reaction O + H$_3^+$
$\to$ OH$^+$ + H$_2$) times 0.33 (the last factor accounts for the
fraction of H$_3$O$^+$ recombinations with electrons forming H$_2$O
\citep{2000A&AS..146..157L}).  The atomic oxygen abundance is also
computed by the steady state equilibrium between formation and
destruction processes: photo-dissociation of gas phase water
(Eq.~(\ref{eq:3}) ) for the formation, and formation of gaseous water
via reaction with H$_3^+$ (Eq.~(\ref{eq:4}) ) and freezing onto the
grains for the destruction of O.  We assume that these
reactions are the dominating processes, and that all the oxygen not
contained in CO or silicates is contained in O, H$_2$O, and H$_2$O ice.

Finally, the H$_3^+$ abundance is computed as follows. H$_3^+$ is
formed by cosmic ray ionization of H$_2$ to form H$_2^+$, followed by
the reaction with H$_2$ to form H$_3^+$.  In the regions where CO
molecules are not frozen onto the grain mantles, H$_3^+$ is destroyed
by the reaction with CO. Elsewhere, the H$_3^+$ abundance is computed
following the model described in \citet{2005submitted}, which
takes into account all three deuterated forms of H$_3^+$ and solves
the chemical composition by considering the reactions between H$^+$,
e$^-$, grains, and H$_3^+$ isotopologues.
\section{Discussion}

\begin{figure}[htbp]
\centerline{\includegraphics[width=9.0cm]{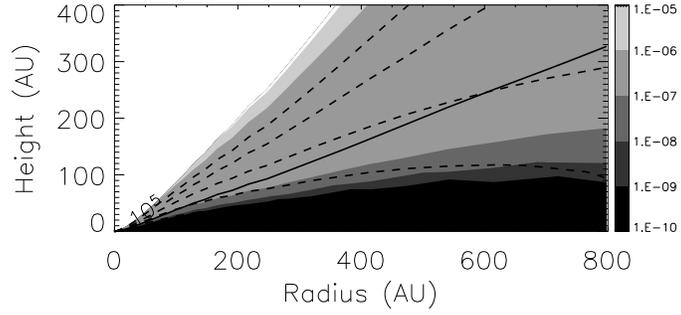}}
\caption{\label{fig:abundance} n(H$_2$O)/n(H$_2$) as a function of
  position in the disk. The solid line shows where the UV optical
  depth measured vertically to the disk plane is unity:
  $\tau_\mathrm{uv}=1$ (see text). The dashed lines show iso density
  contours: from the top to the bottom they are $10^4$, $10^5$, $10^6$
  and $10^7$ cm$^{-3}$.  The white region was excluded from the
    calculation because of the low density there.}
\end{figure}

\begin{figure}[htbp]
\centerline{\includegraphics[width=9.0cm]{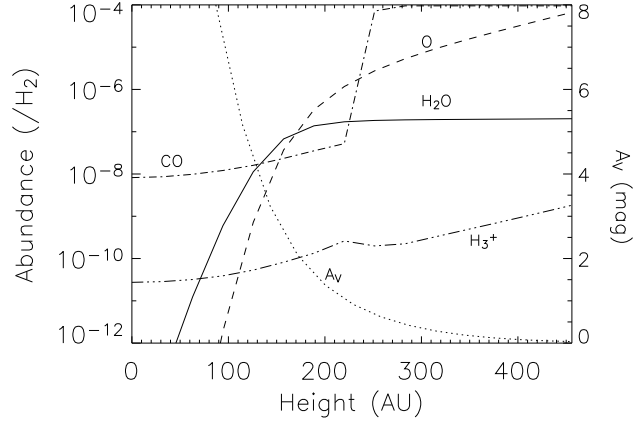}}
\caption{\label{fig:pdr} The abundances and the visual extinction
  A$_v$ (from the disk surface) as a function of the height above
  midplane at 700 AU from the star.  The solid and dashed lines show
  the abundance of H$_2$O and O, respectively.  The dashed-dotted and
  the dotted-dashed lines show the abundance of CO and H$_3^+$.  The
  dotted line indicates the $A_V$.  The most important contribution to
  the H$_2$O column density is produced between $A_V\sim 1$ and
  $\sim4$ mag, and a height between 100 and 250 AU. CO freezes out at
  a height of $\sim230$ AU.  This causes a small increase of the
  H$_3^+$ abundance, since CO (along with O) is the main destroyer of
  H$_3^+$.  The high abundance of CO at very low A$_V$ may be
  artificial because we have ignored CO photo-dissociation.  The water
  ice abundance is not shown because it is almost constant
  $2\times10^{-4}$ throughout the plot.
}
\end{figure}

\begin{figure}[htbp]
\centerline{\includegraphics[width=9.0cm]{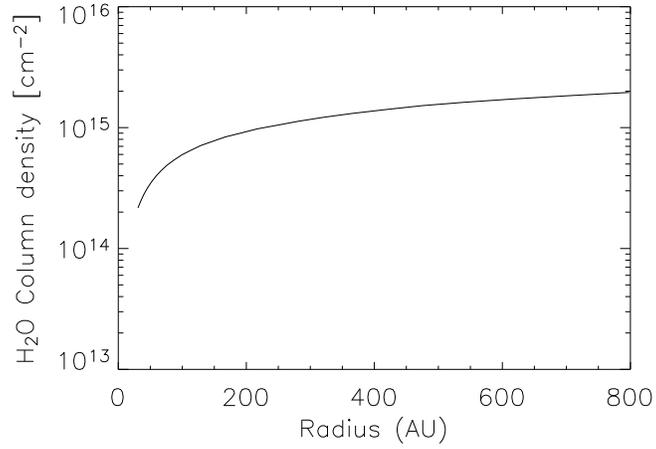}}
\caption{\label{fig:coldens}
  Gas-phase H$_2$O column density measured perpendicular to the disk
  midplane as function of the radius.  In the very inner regions the
  H$_2$O column density decreases because the gas densities at
  A$_v\sim1$ are higher, leading to a greater proportion of H$_2$O
  that is frozen out onto grains as opposed to photo-desorbed. Once the
  grains are hotter than 100 K ($R\la30$AU), water will not freeze-out
  and gas phase abundances will rise again (not shown).}
\end{figure}

Figure \ref{fig:abundance} shows the H$_2$O abundance (with respect to
H$_2$) across the disk, for a standard interstellar UV field
($G_0=1$).  Figure \ref{fig:pdr} shows a vertical cross-section of the
chemical species involved in water formation at a radius of 700 AU.
Despite the relatively large densities in the disk, water vapor has an
abundance of $\sim 3\times 10^{-7}$ in a large fraction of the outer
disk, in the layers just above the midplane, where the A$_{\rm V}$ to
the disk surface is lower than $\sim5$ mag.  By midplane here we mean
the region where more than 2/3 of CO molecules are frozen onto the
grain mantles. This occurs at an height of about 230 AU at a radius of
700 AU, and 100 AU at a radius of 400 AU.  Above this height,
  CO is desorbed thermally\footnote{In our model CO molecules are
  not photo-desorbed from the mantles, although this is plausible,
  depending on the structure of the mantles and where on the grain the
  frozen CO is located.  However, the gas phase CO abundance has only
  a minor influence on the gas phase H$_2$O abundance in our model.}.
The gas phase water abundance peaks near the surface where the FUV
desorbing flux is high and where the H$_3^+$ abundance is high. In
practice, the freeze-out rate of the H$_2$O molecules is larger than
the FUV photo-desorption rate of the grain mantles only at densities
larger than about $10^7$ cm$^{-3}$.  At lower densities, assuming
water is in the grain surfaces, water vapor is formed mostly through
direct photo-desorption from mantles, and, to a lesser extent, by gas
phase reactions occurring among atomic oxygen (a product of H$_2$O
photo-dissociation) and H$_3^+$. In the upper layers, atomic oxygen is
predicted to be very abundant.  The resulting average gas phase H$_2$O
column density of a face-on disk is $1.6\times 10^{15}$ cm$^{-2}$.
Figure \ref{fig:coldens} shows how the column density is distributed
as function of the radius.  The column density is remarkably constant
over most of the disk, although the column does rise slightly with
radius.

The first conclusion of our study is that {\it vapor water can indeed
be found with relatively high abundances ($\sim10^{-7}-10^{-6}$) and
column densities ($\sim 1.6\times 10^{15}$ cm$^{-2}$) in the
protoplanetary disks which surround low luminosity protostars
illuminated by the interstellar radiation field (ISRF), like in the
case of DM Tau}.  This is caused by the photo-desorption of the icy
grain mantles by the average ISRF.  Increasing the FUV field by a
factor of ten and hundred leads to H$_2$O column densities larger by
only factor 1.5 and 2 respectively.  Decreasing the FUV field by a
factor ten results in decreasing the H$_2$O column density by a factor
a bit more than a factor 2.  Therefore, the H$_2$O column density is
not sensitive to the addition of a possible FUV field from the central
source or a nearby hot star (see Hollenbach et al. 2005 for details
and an analytical proof of the insensitivity to G$_o$).  The
fundamental reason for this insensitivity is that in higher fields,
the increased photo-desorption of the ice is balanced by the increased
photo-dissociation of gas phase H$_2$O.

As discussed above, the value of the photo-desorption yield, $Y$, is
only constrained within a factor roughly three to ten by
laboratory experiments.  This uncertainty causes an uncertainty in the
predicted H$_2$O column density by the same factor.  Another
important, and poorly known parameter that enters in these
computations is the dust-to-gas ratio.
In the standard case we assumed the canonical mass dust-to-gas ratio
equal to 1\%.  If this value is increased by a factor ten (which means
decreasing by a factor 10 the gas density in the plot of
Fig.~\ref{fig:structure}), the H$_2$O column density increases by a
factor $\sim1.2$. On the contrary, a decrease by a factor ten of the
dust-to-gas ratio leads to a decrease by a factor 1.2 of the H$_2$O
column density.  This is due to the fact that the UV optical
depth is determined by the dust distribution only and independent of
the gas density.  As long as photo-desorption, photo-dissociation and
freeze-out are the dominant processes, the derived densities and
column densities (as opposed to the abundances) of H$_2$O are
independent of the gas density.  The small changes are due to regions
in which the gas formation rout of H$_2$O is important.  Therefore,
also a drastic change in the dust-to-gas ratio does not cause much of
a change in the H$_2$O vapor column density.  Finally, the results
will also depend on the grain size distribution.  If much smaller
grains than 0.1$\mu$ dominate the grain surface area, the balance
between the different formation and destruction rates will be shifted.
This question will be discussed in a followup paper.

In conclusion, we predict a H$_2$O column density in protostellar
disks similar to that surrounding DM Tau of a few times $10^{15}$
cm$^{-2}$, rather insensitive on the external FUV field and/or the
dust-to-gas ratio in the disk.


\citet{HDO-DMTAU} reported the detection of HDO in DM Tau, with an
observed column density of $\sim 1.6 \times 10^{13}$ cm$^{-2}$.  Based
on the present theoretical model, the H$_2$O column density is
predicted to be $\sim 2\times10^{15}$ cm$^{-2}$, within about a
factor of a few, given the uncertainties in the yield, FUV field and
the dust-to-gas ratio.  This implies a HDO/H$_2$O ratio equal to about
1\%.  This is consistent with estimates of the HDO/H$_2$O ratio in
embedded low mass protostars.  \citet{2005A&A...431..547P} measured a
HDO/H$_2$O ratio equal to 3\% in the sublimated ices surrounding
IRAS16293-2422, and a much lower ($\leq$0.2\%) value in the outer
envelope where the gas phase chemistry dominates the HDO and H$_2$O
formation. Note that the large HDO/H$_2$O ratio of 3\% refers only to
the sublimated ices, which are responsible for the water abundance
$3\times10^{-6}$ in this object \citep{2000A&A...355.1129C}.
Very likely,
the sublimated ices are only the ``last coating'' of the mantles where
the molecular deuteration is the largest, due to the history of ice
formation. In agreement with this interpretation, NIR observations of
HDO ice have, so far, failed to detect it, at a level of
HDO/H$_2$O$\leq2$\% 
\citep{Pariseetal2003a}
Therefore, only a small fraction of elemental deuterium is iced on the
mantles. The HDO/H$_2$O ratio in DM Tau, equal to $\sim1$\%, is just
slightly lower than the value measured in IRAS16293-2422 and
attributed to sublimated ices.  If the values were indeed different,
this might point to reprocessing of the ices in the disk - however,
for this an actual measurement of the vapor water in DM Tau is
required.  The measurement of the H$_2$O column density would also
allow to measure the photo-desorption yield.  In addition, measures of
the HDO/H$_2$O ratio in protoplanetary disks will probe the history
of the ice formation, with important implications for the chemical and
isotopic composition of Solar System objects.  Based on our model, we
predict the ground state transition of H$_2$O at 557 GHz to be in
absorption similar to the HDO ground transition at 464 GHz.  The line
is predicted to be highly optically thick, with $\tau\sim500$.
Unfortunately, neither the currently available ODIN nor the planned
HSO satellites will be able to detect the continuum of DM Tau, because
of the large beam dilution of the signal from the disk.

\section{Conclusions}

The results reported in this \emph{Letter} lead to the following
important conclusions:\\
1. An external UV radiation field as weak as the average Habing field
(G$_0$=1) can produce a layer of H$_2$O on top of a protoplanetary
disk (like that surrounding DM Tau) with a
column of a few times 10$^{15}$cm$^{-2}$.\\
2. This column density is insensitive to external parameters like the
strength of the radiation field: Variations in G$_0$ by two orders of
magnitude only change
N(H$_2$O) by a factor of 2.\\
3. Similarly, changing the dust-to-gas ratio (i.e. the gas density at
a given dust density) has only weak influence
on the derived column densities.\\
4. Because of the weak dependence on parameters, the measurement of
the H$_2$O column density in protoplanetary disks is an excellent
tool to measure the
photo-desorption yield $Y$.\\
5. Combined with previous observations of HDO \citep{HDO-DMTAU}, we
predict an HDO/H$_2$O ratio in the outer disk of DM Tau of 0.01, with
an uncertainty of
about a factor of a few.\\
6. We predict that the H$_2$O ground state line at 557 GHz is
undetectable in DM Tau with ODIN and/or HSO-HIFI
observations.\\
Finally, we remark that other molecules trapped in the water ices
could also be released into the gas phase, keeping ``alive'' the gas
chemistry in the layers above the midplane of the outer disk.

\begin{acknowledgements}
  CD and CC acknowledge Travel support through the Dutch/French van
  Gogh program, project VGP 78-387 and the hospitality at NASA Ames
  during the development of this work, during which CD received Travel
  funds from the University of California, Berkeley.  DH acknowledges
  support from the NASA Astrophysical Theory, Dynamics and Origins of
  Solar Systems.  This work has benefited from research funding from
  the European Community's Sixth Framework Programme. We thank the
  referee Juri Aikawa for a detailed report.
\end{acknowledgements}

\clearpage

\bibliographystyle{apj}

\clearpage

\clearpage

\clearpage

\clearpage

\end{document}